\documentstyle[]{article}
\begin{document}
\title{ Quantum  Fields in Non-Static Background: A 
Histories Perspective} 
\author {C. Anastopoulos \thanks{E-mail:charis@physics.umd.edu}  \\ {\small
Departament de Fisica Fonamental, Universitat de Barcelona} \\
{\small  Av. Diagonal 647, 08028 Barcelona, Spain} \\ \\ 
  {\small and} \\ {\small Department of Physics,
 University of Maryland,} 
{\small College Park, MD20742, USA} \\
{\small (current address)}}
\maketitle
%\begin{center}
%{\large{\bf Quantum Fields in Non - Static Background: a Histories Perspective% }} \vskip 1.2cm C. Anastopoulos\footnote{e-mail:
%charis@physics.umd.edu } \vskip 0.4cm {\it 
%Departament de Fisica Fonamental, Universitat de Barcelona \\
%  Av. Diagonal 647, 08028 Barcelona, Spain  \\ 
% and \\  Department of Physics,
% University of Maryland \\
% College Park, MD20742, USA \\
% (current address)}
%\end{center}
%\vskip 0.8cm
\begin{abstract}
 For a quantum field living on a non - static spacetime no instantaneous
 Hamiltonian is definable, for this generically necessitates a choice
 of inequivalent representation of the canonical commutation relations at
 each instant of time. This fact suggests a description 
in terms of time - dependent Hilbert spaces, a concept that  fits naturally in
a (consistent) histories framework.  Our primary tool for the construction of the quantum 
 theory in a continuous -time histories format is the recently developed formalism based on 
the notion of
 the history group . This we employ to   study   a model system
 involving a 1+1 scalar field in a cavity with
moving boundaries.
 The instantaneous (smeared)  Hamiltonian   and a decoherence functional
are then rigorously defined so that finite values for the time - averaged
particle creation rate are obtainable through the study of energy histories.
We also construct the Schwinger - Keldysh closed- time - path  generating
functional as a ``Fourier transform'' of the decoherence functional and
evaluate the corresponding  n - point functions.
\end{abstract}

\renewcommand {\thesection}{\Roman{section}} 
 \renewcommand {\theequation}{\thesection. \arabic{equation}}
\let \ssection = \section
\renewcommand{\section}{\setcounter{equation}{0} \ssection}
\pagebreak

\section{Introduction}
The consistent histories approach \cite{Gri,Omn,GeHa,Har} was mainly
devised as an alternative point of view to quantum phenomena, providing
 a more convenient language for the treatment of individual, closed quantum
 mechanical systems. While its physical predictions exactly agree with the
 ones of standard quantum mechanics (arguably even for the case of the
 paradoxes connected to the multiplicity of the consistent sets),  its
 internal structure is somehow distinct. While standard quantum mechanics (in
 its Heisenberg version) incorporates kinematics through Hilbert space
 operators, 
dynamics through a Hamiltonian and initial conditions (probability
assignment) through a state (density matrix), in the history theory histories 
(or history propositions) provide the kinematics , with   state and dynamics
being encoded in a new object: the decoherence functional. 
\par
This is a complex valued function $d(\alpha, \alpha')$ of pairs of histories,
whose role is the assignement of probabilities. If for any set of histories,
the decoherence condition 
\begin{equation}
d( \alpha, \alpha') = 0
\end{equation}
is satisfied for $\alpha \neq \alpha'$, then a probability measure exists in this set
given by $p(\alpha) = d(\alpha, \alpha)$. 
\par
A mathematically elegant  formalism for histories has been developed by Isham
and collaborators \cite{Ish12,ILS} . In this formulation
histories can be
identified with projection operators on a Hilbert space. In standard systems
this is constructed  from the tensor product of the single - time
Hilbert spaces, that
characterise the canonical theory. Besides providing a characterisation of consistent histories
as temporal quantum logic,
this formalism highlights the   its similarity to their  closest
classical analogue:  stochastic processes.
An important feature of this construction is that these single-time Hilbert
spaces need not be isomorphic
 (or carry isomorphic structures such as unitarily equivalent group
 representations ) to each other.
\par
This seems  particularly suited for the study of quantum field theories in
non-static background, for the following reason : The  Hilbert space 
that defines the quantum  theory corresponding to a particular classical
system is constructed from the study of the representations of the canonical
  group (for the general scheme see reference  \cite{GGT}). For linear
  systems  this is the familiar Weyl group, the Lie algebra of which is
  defined by 
the canonical commutation relations.
 When considering fields we have to deal with   an infinite dimensional Lie
 group, which  will  admit many unitarily inequivalent representations. 
 The natural way to proceed would be to select a representation, 
in which   the Hamiltonian can  defined as a concrete self -adjoint operator.
When this is attempted for fields in a non-static background, one realises
 that at different times one has 
to admit unitarily inequivalent  representations of the canonical commutation
relations. This implies the non -existence of an instantaneous Hamiltonian.
 Rather than abandoning the definition of a Hamiltonian, 
the histories approach provides a possibility of welding these
representations together in order to construct a well- defined {\it finite}
quantum  theory describing such systems.
\par
Two  recent developements provide insight necessary for dealing with this
case. In a series of papers Isham {\it et al} analysed the kinematical
structure 
of the Hilbert space describing continuous - time histories
\cite{IL,ILSS}. The main ingredient has been the history group, the analogue
of the  canonical group  in the histories context. Its Lie algebra for the
case of a particle at a line is
\begin{equation}
[x_t,p_{t'}] = i \hbar \delta(t,t')
\end{equation}
The time index $t$ does not here refer to the dynamics of the system, as
generated by a Hamiltonian, but is an index labeling the instant of time at
which a proposition (for instance corresponding to the generators) is
asserted. 
Equation (1.2) is formally similar to the canonical algebra for an 1+1 field
theory and as such admits many unitarily inequivalent representations. A
guiding
 principle for a selection of a representation has been the definability of
 an instantaneous Hamiltonian \cite{ILSS} . Accepting continuous time implies
 that  
all history propositions are  about quantities smeared (averaged) in time.
This again suggests that if we demand the existence of a smeared,
instantaneous 
 Hamiltonian,  we might be able to obtain   a unique representation of the
 history algebra corresponding to a quantum field in non-static 
background.
\par
This we shall show that can be relatively straightforwardly achieved through
a simple generalisation of the results of reference \cite{ILSS}. But
 then we should need a guiding principle for the construction of the
 decoherence functional, since the corresponding canonical theory is not well
 defined . This has come from a recent result by Savvidou: the discernment of
 two laws of time transformations (and corresponding time parameters) 
 in the  backbone of the structure 
of history theories \cite{Sav}.  One parameter is associated with the
background temporal structure and describes how one moves from one single -
time
 Hilbert space to another (Schr\"odinger time). The other  incorporates the
 effects of the actual dynamics (Heisenberg time). Taking this as a
 fundamental property that ought 
to be reflected in all objects of our theory , we have been able to  expand
on a previous partial result \cite{IL}  and identify three pieces out of
 which a physical decoherence functional is constructed. These pieces we
 call: 
the Schr\"odinger operator, the Heisenberg operator and the boundary
operator. They correspond respectively to the aforementioned  times and the
 initial state. 
\par
This result enables us then to write a {\it finite} decoherence functional
for a model case we study in this paper: an 1+1 field in a cavity with 
moving boundaries. This is sufficiently general to prove the main point: that
a history theory based on the physical principle of the existence of 
an instantaneous Hamiltonian rigorously exists and can provide  finite values
for the  probabilities assigned to  histories \footnote{One can view these
results as a direct consequence of 
the existence of two notions of time transformation in history theory
compared to
 the unique one of canonical quantum mechanics. In the canonical theory the
 non - homogeneity of time transformations reflects itself in the 
non - existence of the corresponding generator. In the history version it is
only the generator of {\it kinematical} time transformations, related 
to the spacetime causal structure (the Liouville operator of reference
\cite{Sav}) that  need not exist, while the other (the Hamiltonian) does
exist, 
and it is the one that determines the representation space of the theory.}.
In addition, we establish that the Schwinger - Keldysh 
closed - time  - path (CTP) generating functional  \cite{CTP} is equal to the
decoherence functional evaluated at a pairs of elements of the history 
group. Hence we are able to construct n - point functions and  get into
contact with the results of the more familiar canonical treatment. 
\par
 The generalisation of these results to general spacetimes is technically
 straightforward ( by demanding the existence of the Ashtekar-Magnon
 Hamiltonian
 \cite{AM}), but for one thing. Our theory should be independent of the
 choice of the time variable employed in the definition of the Hamiltonian.
 This means that changes of foliation should be generically implemented by a
 unitary operator on the history Hilbert space \cite{Schr}. Such a proof of 
unitary implementation for the general case is quite more demanding and
necessitates a different set of techniques from the ones we employ in this
paper.
We therefore  defer it  to a 
future work.
Here we shall restrict ourselves to the convenient choice of the
background Minkowski time, which in any case is relevant for the
 discussion of the time - dependent Casimir effect.

\section{ Structure of decoherence functional} 
 All physical histories can be represented by elements of a lattice of
 propositions \cite{Ish12} and in the familiar case of standard quantum
 mechanics they are realised by 
projection operators on a Hilbert space ${\cal F}$ which is the tensor
product of the single time Hilbert spaces
\begin{equation}
{\cal F} = H_{t_1} \otimes H_{t_2} \otimes \ldots \otimes H_{t_n}
\end{equation}
\par
In \cite{ILS} the most general form of a decoherence functional satisfying
the relevant axioms has been constructed. It is in one to one 
correspondence with particular class of operators $X$ that act on ${\cal F}
\otimes {\cal F}$. Explicitly
\begin{equation}
d(\alpha, \alpha') = Tr_{{\cal F} \otimes {\cal F}} \left( P_{\alpha} \otimes
P_{\alpha '} X \right)
\end{equation}
This is an important result, enabling a mathematical classification of
decoherence functionals, 
but what would be more interesting for physical aplications is the
construction of the decoherence functional in terms of operators acting
solely 
on ${\cal F}$. ( This has been done for a special case in reference
\cite{IL}).  The reason for that is the possibility of having  a physical 
interpretation and understanding for such  objects. This would  enable the
construction of  the decoherence functional even when not having the
reliable guide of a corresponding canonical theory. This is what we shall
undertake in this section. 
\par
Our starting point is the standard form of the time symmetric decoherence
functional \cite{Har} 
\begin{equation}
d(\alpha,\alpha') = Tr_H (C_{\alpha'}^{\dagger} \rho_f C_{\alpha}
\rho_0)/Tr(\rho_f \rho_0)
\end{equation}
with $C_{\alpha} = P_{\alpha_n}(t_n) \ldots P_{\alpha_1}(t_1)$ in terms of
the Heisenberg picture projectors and the trace is performed within 
the single-time Hilbert space $H$. Recall that the standard form can be
obtained by setting $\rho_f = 1$.
\par
It is easy to verify   \cite{ILS} that the above expression can be written
in the form 
\begin{equation}
Tr_{\otimes_{i=1}^4 H_i} \left(C_{\alpha'}^{\dagger} \otimes \rho_f \otimes
C_{\alpha} \otimes \rho_0 S_4 \right)/ Tr(\rho_f \rho_0)
\end{equation}
where $S_{4}$ is an operator acting on $H \otimes H \otimes H \otimes H$  as
\begin{equation}
S_{4} ( |v_1 \rangle \otimes |v_2 \rangle \otimes |v_3 \rangle \otimes |v_4
\rangle) =
|v_2 \rangle \otimes |v_3 \rangle \otimes |v_4 \rangle \otimes |v_1 \rangle
 \end{equation}
Tracing independently over the second and fourth Hilbert space in (2.4) we
get the expression
$Tr_{H \otimes H} (C_{\alpha'}^{\dagger} \otimes C_{\alpha} Z)$ where 
\begin{equation}
Z = Tr_{H_2\otimes H_4} (1 \otimes \rho_f \otimes 1 \otimes \rho_0)
S_{4}/Tr(\rho_f \rho_i)
\end{equation}
in an obvious notation.
The matrix elements of $Z$ in an  orthonormal basis $|i \rangle \otimes |
j\rangle $ of $H \otimes H$ are easily computed
\begin{equation}
 \langle m l |Z| i j \rangle = (\rho_f)_{mj} (\rho_0)_{ki}/Tr(\rho_f \rho_0)
\end{equation}
This can be written  in the form
\begin{equation}
\langle m l |Z| i j \rangle = \sum_{rs} \left[ (\rho_f^{1/2})_{mr}
(\rho_0^{1/2})_{si}\right] \left[ (\rho_f^{1/2})_{rj} (\rho_0^{1/2})_{ls}
\right]/ 
Tr(\rho_f \rho_0)
\end{equation}
hence
\begin{equation}
Z = \sum_{rs} A^{(r,s)}\otimes (A^{\dagger})^{(r,s)}
\end{equation}
with $A^{(rs)}$ operators on $H$ with matrix elements
\begin{equation}
\langle i | A^{(rs)} | j \rangle = \left[(\rho_f^{1/2})_{ir}
(\rho_0^{1/2})_{sj}\right]/\left( Tr(\rho_f \rho_o) \right)^{1/2} 
\end{equation}
Since the history operators are trace-class and the $A$ 's bounded, the
decoherence functional can  be written
\begin{equation}
d(\alpha,\alpha') =  \sum_{rs} Tr_H \left( C_{\alpha'}^{\dagger} A^{(rs)}
\right) Tr_H \left( C_{\alpha} A^{\dagger (rs) } \right) 
\end{equation}
In the above expression the decoherence functional has separated in different
traces the contribution of each of the pair of histories. If in each of 
those traces we employ the technique we used to derive (2.4) we obtain
\begin{eqnarray}
d(\alpha, \alpha') = \sum_{rs} Tr_{{\cal F}} \left[ {\cal U}^{\dagger}P_{\alpha}{\cal
U} {\cal A}^{\dagger}{ }^{(rs)}{\cal RSR} \right] 
 Tr_{{\cal F}} \left[ {\cal U}^{\dagger} P_{\alpha'} {\cal U} {\cal A}^{(rs)} 
{\cal S} \right]
\end{eqnarray}
where 
\begin{eqnarray}  
P_{\alpha} = P_{\alpha_1} \otimes \ldots \otimes P_{\alpha_n} 
\end{eqnarray}
is the projector on ${\cal F}$ corresponding to the history proposition
$\alpha$ and similarly for $P_{\alpha'}$. Also
\begin{eqnarray}
{\cal U} = U_1 \otimes \ldots U_n = e^{-iHt_1} \otimes \ldots \otimes
e^{-iHt_n} \\
{\cal A}^{(rs)} =  A^{(rs)} \otimes 1 \otimes \ldots \otimes 1
\\
\end{eqnarray}
and the operators ${\cal S}$ and ${\cal R}$ are defined in terms of their
action
\begin{eqnarray}
{\cal S} ( |v_1 \rangle \otimes \ldots \otimes | v_n \rangle ) = |v_2 \rangle 
\otimes | v_3 \rangle \otimes \ldots \otimes |v_1 \rangle  \\
{\cal R}( |v_1 \rangle \otimes \ldots \otimes | v_n \rangle ) = |v_n \rangle 
\otimes | v_{n-1} \rangle \otimes \ldots \otimes |v_1 \rangle  
\end{eqnarray}
\par
Before discussing the physical significance of these operators let us recast
our expressions in a more elegant and suggestive form. Let us
 by $\partial_- {\cal F}$ and $ \partial_+ {\cal F}$ denote the past and
 future ``boundary'' of ${\cal F}$, that is the Hilbert spaces $H_{t_1}$ 
and $H_{t_n}$ in the case of discrete histories we have considered in this
section. If ${\cal H}$ is the space of continuous linear maps
 $\partial_- {\cal F} \rightarrow \partial_+ {\cal F}$ then ${\cal A}$ can be
 viewed as a linear map from ${\cal F}$ to ${\cal H}$. Checking 
that ${\cal RSR} = {\cal S}^{\dagger}$ and assuming the initial and final
times to be the ones at which $\rho_0$ and $\rho_f$ are defined our 
expression (2.12) reads
\begin{equation}
d(\alpha, \alpha') = Tr_{{\cal H}} \left[ Tr_{{\cal F}} \left( {\cal U}^{\dagger}
P_{\alpha } {\cal U} ({\cal SA})^{\dagger}\right)
 Tr_{{\cal F}} \left( {\cal U}^{\dagger} P_{\alpha'}  {\cal U} ({\cal SA})
 \right) \right]
\end{equation} 
hence the operator $X$ of (2.2) is given by
\begin{equation}
X = Tr_{{\cal H}} \left(  {\cal U} {\cal A}^{\dagger} {\cal
S}^{\dagger} {\cal U}^{\dagger} \otimes
{\cal USAU}^{\dagger} \right)
\end{equation}
\subsection{ Interpretation}
It is clear from the discussion above that three are the important
ingredients entering in the construction of the decoherence functional: \\
\\
1. The unitary operator ${\cal U}$ in which the contribution of the 
dynamics in the evaluation of probabilities 
are contained. Its action is to provide the weight in the probabilities due
to time evolution, or, rather heuristically, to turn the projection
 operators into Heisenberg picture ones. Note that this operator can be
 similarly defined  even when the Hamiltonian is time dependent.
\\
\\
2. The operator ${\cal S}$. As can be seen from its definition it encodes the
temporal structure (in the sense of partial ordering) of the history 
theory. In the standard (discrete time) case we have examined, it can be
readily verified to be unitary, but it seems reasonable that this condition 
could be relaxed in certain generalisations  \footnote{What we have in mind
is spacetimes that are not globally hyperbolic, as possibly involving
topology change.}. A sufficient condition for the finiteness of the traces in
(2.19) is 
that ${\cal S}$ is bounded, which follows trivially when being unitary. Note
that also by its definition
$Tr {\cal S} = 1$.
\\
\\
3. The linear map ${\cal A}$. It essentially contains the contribution of the
initial and final states, that is the weight given to probabilities
 by the particulary boundary conditions. In the standard case it is a
 trace-class operator but in general (as in our particular examples later) we 
can dispense with even its (strong)continuity. Given our previous assumptions
for ${\cal U}$ and ${\cal S}$ a sufficient condition on ${\cal A}$ for the 
finiteness of the traces is the
\\ \\
 {\bf Weak continuity condition:} \\
1.$Tr_{{\cal H}} \left( {\cal A}^{\dagger} {\cal A} \right) : {\cal F }
\otimes \bar{{\cal F}}
 \otimes {\cal F} \otimes \bar{{\cal F}} \rightarrow {\bf C}$ is  continuous.
\\ 
2. For any $|\phi_1 \rangle, | \phi_2 \rangle \in {\cal H}$,
$\langle \phi_1 | {\cal A} | \phi_2 \rangle : {\cal F} \rightarrow {\bf C}$
is continuous. \\ \\ 
For purposes of  easy reference we shall henceforward call ${\cal U}$  the
Heisenberg operator, ${\cal S}$  the Schr\"odinger operator
and ${\cal A}$ the boundary operator.
\par
Casting the decoherence functional in the form (2.19) can be seen as a step
of departure for constructing generalised history theories. A non-trivial
 generalisation is when the single-time Hilbert spaces are not the same . The
 change is then included in the operator ${\cal S}$ which contains 
information about the welding of the unequal time Hilbert spaces together in
the history Hilbert space ${\cal F}$. If there exist (physically justifiable)
 identification maps between the Hilbert spaces $H_{t_i}$ (to keep full
 generality this does not have to be  structure preserving ), i.e. a family
 of
 maps
\begin{equation}
I(t_i,t_j): H_{t_j} \rightarrow H_{t_i}
\end{equation}
then the Schr\"odinger operator can be defined as
\begin{equation}
{\cal S} (|v_1 \rangle \otimes |v_2 \rangle \otimes \ldots \otimes |v_n
\rangle) = I(t_2,t_1) |v_1 \rangle \otimes I(t_3,t_2) |v_2 \rangle \otimes 
\ldots I(t_1,t_n) |v_n  \rangle
\end{equation}
\par
Finally, we should remark that the behaviour of ${\cal S}$ at the boundaries
would results in most situations in its intricate mixing with the boundary 
operator ${\cal A}$.

\section{Scalar field in time dependent spacetime }
The identification of the operator structure within the decoherence
functional carried out in the previous section, enables us to proceed in the 
construction of history theories, in which there is a time-dependence on the
single time Hilbert spaces.  Our motivation is the study of field theories
 in non - static background; hence  in the rest of the paper we shall
 undertake an examination of a  simple model, employing techniques developed
 for
 the study of continuous histories. 
\subsection{The history algebra}
The system under study is  a massless scalar field in a cavity with
time-varying size given by the positive function $L(t)$ (this has to be
assumed 
to have at least continuous first derivative). On the edges of the cavity
Dirichlet boundary conditions are to be assumed. The time parameter with 
respect to which the history theory is defined is the ``Minkowski'' time.
\par
Our natural choice for the history algebra is
\begin{equation}
[\phi_t(x), \pi_{t'}(x')] = i \frac{1}{L(t)} \delta(t, t') \delta_{I}(x,x')
\end{equation}
where $t$,$t'$ lie in ${\bf R}$, $x$,$x'$ in $I = [0,1]$, $\delta_{I}$
denotes the delta function as defined on  $I $ and Dirichlet boundary
conditions are
 assumed for the fields \footnote{ The $\delta$ function is  represented as
 $\sum_{n=1}^{\infty} \sin n \pi x $. The presence of the $1/L(t)$ in the 
right hand side, is due to the fact that the proper integration measure is
$L(t) dx$.}. More precisely one should use the smeared fields 
\begin{eqnarray}
\phi(f) = \int dt L(t)\int_0^1 dx \phi_t(x) f(t,x) \\ 
\pi(g) = \int dt L(t) \int_0^1 dx \phi_t(x) g(t,x)
\end{eqnarray}
where $f$ and $g$ are elements of the vector space $L^2_R(R) \otimes
L_R^2(I)_D$ where $D$ stands for the imposition of the Dirichlet boundary
condition
\begin{eqnarray}
f(0,t) = f(1,t) = 0
\end{eqnarray}
This way we have
\begin{equation}
[\phi(f), \pi(g)] = i \int dt L(t) \int_0^1 dx f(t,x) g(t,x)
\end{equation}
It is more convenient to express the fields in terms of their Fourier
transforms
\begin{eqnarray}
\phi_t(x) = L^{-1/2}(t) \sum_{n=1}^{\infty} q_t(n) \sin n \pi x \\
\pi_t(x) = L^{-1/2}(t) \sum_{n=1}^{\infty} p_t(n) \sin n \pi x
\end{eqnarray}
with respect to which 
\begin{equation}
[q_t(n),p_{t'}(m) ] = i \delta_{nm} \delta(t,t')
\end{equation}
These can be smeared by elements of $L_R^2(R)_D$, so that $q_n(f) = \int dt
q_t(n) f(t)$. Hence
\begin{equation}
[q_n(f),p_m(g)] = i \delta_{mn} \int dt f(t) g(t)
\end{equation}
 and the underlying vector space of the history algebra is simply
 $L^2_R(R)_D \otimes l^2_R$.

\subsection{ The representation space}
Our next task is to find a representation of the history algebra (3.1) in a
Hilbert space ${\cal F}$, together with an isometry from the continuous 
tensor product $\otimes_{t \in R} {\cal H}_t$, where ${\cal H}_t$ is the
Hilbert space on which the {\it canonical algebra} at time $t$ is
represented. 
This Hilbert space supposedly contains the projections having information
about the properties of of the system at time $t$.

\subsubsection{Single-time Hilbert spaces ${\cal H}_t$}
The canonical algebra at time $t$ reads
\begin{equation}
[\phi(x), \pi(x')] = \frac{i}{L(t)}\delta_I(x-x')
\end{equation}
Now ${\cal H}_t$ has naturally the structure of an exponential Hilbert space:
$ {\cal H}_t = e^{V_t} = \oplus_{n=0}^{\infty} (V_t)^n_S$ where $V_t$ is the
space of complex valued functions on $[0,1]$ satisfying Dirichlet boundary 
conditions and with inner product given by
\begin{equation}
(z_1,z_2)_t = L(t) \int_0^1 dx z_1^*(x) z_2(x)
\end{equation}
It is well known that  ${\cal H}_t$ is spanned by an overcomplete set of
states (unnormalised coherent states ) 
$| \exp z \rangle = \oplus_{n=0}^{\infty} \otimes_n z $, ($ z \in V_t$) with
inner product
\begin{eqnarray}
\langle \exp z | \exp w \rangle_t = \exp \left( L(t)\int_0^1 dx z^*(x) w(x)
\right) 
\end{eqnarray}
Equivalently using the Fourier transform
\begin{eqnarray}
f(x) = L^{-1/2} \sum_{n=1}^{\infty} z_n \sin n \pi x
\end{eqnarray}
 one characterises $V_t$ as $L^2(R)_D \otimes l^2$. 
\subsubsection{ The history Hilbert space ${\cal F}$}
The fact that ${\cal H}_t$ can be writen as an exponential Hilbert space
enables us to employ the analysis of \cite{IL} for the 
construction of ${\cal F}$. The important identity that carries on into our
case is
\begin{eqnarray}
\langle \otimes_t \exp z _t | \otimes_t \exp w _t \rangle_{\otimes_t e^{V_t}}
= \exp \left( \int dt ( z_t|w_t)_{V_t} \right) \nonumber \\
= \exp \left( \int dt  \sum_{n=1}^{\infty} (z^*_n)_t (w_n)_t \right)
\end{eqnarray}
where $z$ and $w$ stand for elements of $l^2$. 
This implies straightforwardly the isomorphisms
\begin{eqnarray}
 \otimes_t \exp V_t &\simeq& \exp \int^{\oplus} V_t \simeq \exp {\cal E} 
\nonumber \\
 \otimes_t | \exp z_t \rangle &\rightarrow& | \exp \int^{\oplus} z_t dt
 \rangle
\rightarrow | \exp z(.) \rangle
\end{eqnarray}
where ${\cal E}$ is the space $ L^2(R) \otimes l^2$ (i.e. some
complexification of the test function space in the history algebra) with
inner product
\begin{equation}
\langle z | w \rangle_{{\cal E}} = \int dt  \sum_{n=1}^{\infty} z^*_n(t)
w_n(t)  
\end{equation}
So we conclude that
\begin{eqnarray}
{\cal F} = \otimes_t {\cal H}_t \simeq \exp {\cal E}
\end{eqnarray}

\subsection{ Operators on  F}
Having found the structure of ${\cal F}$ does not yet mean that we have
identified the representation of the canonical group. In each Fock  space 
construction, creation and annihilation operators are naturally defined (see
for instance \cite{Ber}), but we have still to determine how the
 generators of the history group are written as linear combinations of them
 (equivalent to choosing a complex structure on $L^2_R(R) \otimes l^2_R$,
 the space of smearing functions of the history group \cite{Wald}).

\subsubsection{ The Hamiltonian operator}
In standard quantum field theory in Minkowski spacetime this is achieved by
postulating that the state corresponding to the vector $| \exp 0 \rangle$
is invariant under the action of the Poincar\'e group as is represented on
the Fock space \cite{SW}, or equivalently that it is the lowest eigenvalue
 of the field 's Hamiltonian. In generalising to field theory in curved
 spacetime one demands that this state is invariant under the spacetime group 
of isometries. In general, it is essential that this group has a generator
that corresponds to  a timelike Killing vector field , in order that 
time evolution can be unitarily implemented.
\par
This interconnection  between the choice of the representation and the
existence of a Hamiltonian type of operator has been exploited for the case
 of the history group in \cite{ILSS}, in order both to select a Fock space
 representation among the ones considered 
 in \cite{IL} and construct an operator, the spectral family of which can
 naturally be said to correspond to history propositions about energy.
\par
In our  system, the absence of a time translation symmetry would disallow the
definition of a Hamiltonian in a canonical framework 
(mainly for the inability to select a representation of the CCR), but there
is nothing forbidding the introduction of a smeared Hamiltonian
in  the corresponding history theory, along the lines of \cite{ILSS}.
 This would  correspond to the spatially integrated $00 $ component of the
 energy momentum tensor and should act as  an operator governing the
 evolution within a single -  time Hilbert space. 
\par
Keeping these remarks in  mind, we can now proceed to the introduction of the
Hamiltonian $H_t$. Formally, this should be
\begin{eqnarray}
H_t = L(t)\int_0^1 dx \frac{1}{2} \left( \pi_t(x)^2 + ( L^{-1}(t) \partial_x
\phi_t)
(x) ^2 \right) \nonumber \\
= \frac{1}{2} \sum_{n = 1}^{\infty}  \left[ p_t(n)^2 + \left(\frac{n
\pi}{L(t)} \right)^2 q_t(n)^2 \right]
\end{eqnarray} 
 One can then proceed to define suitable creation and annihilation operators
\begin{equation}
a_t(n) = \left( \frac{n \pi}{2 L(t)} \right)^{1/2} q_t(n) + i \left(
\frac{L(t)}{ 2 n \pi } \right)^{1/2} p_t(n)
\end{equation}
thereby making concrete the choice of our representation, and inheriting
${\cal E}$ with a particular complex structure \cite{Wald}.
\par
With respect to these, the Hamiltonian is written
\begin{equation}
H_t = \sum_{n=1}^{\infty} \frac{n \pi}{L(t)} a^{\dagger}_t(n) a_t(n) +
E^{0}(t)
\end{equation}
The terms $E^{vac}(t)$ stands for $\sum_{n=1}^{\infty} \frac{n \pi}{2 L(t)}$
that corresponds to vacuum ``energy''. This is formally divergent and cannot
be physically normal ordered away. Indeed with proper regularisation
 this is exactly the Casimir energy of the field between the plates. In our
 case it is rather straightforward to calculate it using the standard 
point - splitting methods \cite{BD}, with the boundary condition that $E^0 =
0$ for $L(t) \rightarrow \infty$ pointwise.  Indeed, the $t$- dependence 
has no consequence in the computational details. The result is simply
\begin{equation}
E^{0}_{ren}(t) = - \frac{ \pi}{3 L(t)}
\end{equation}
In order to avoid a natural  misunderstanding , we should stress here that
$E^{0}$ is {\it not} the physical vacuum energy of the fields, i.e. it is not 
the value of energy one would read when, say, measuring the force on the
plates. It is rather the lowest {\it possible} energy the field can have at 
any time, and is not expected to be ever naturally realised at all times even
when the system starts in a vacuum state (except of course for the
 trivial case that $L(t)$ is constant). The physical energy at any time is to
 be determined by a proper examining of the energy histories and is 
expected to be a sum of $E^{0}$ with energy due to the  ``particle
creation''. 
\par
The spectrum of $H_t$ is easily identified from  (3.20). The vacuum $|0
\rangle$ (identified with the vector $|\exp  0 \rangle$)  satisfies
\begin{equation}
a_t(n) |0 \rangle = 0
\end{equation}
and corresponding ``many particle'' states
\begin{equation}
| n_1,t_1 ; \ldots ; n_s t_s \rangle = a^{\dagger}_{t_1}(n_1) \ldots
a^{\dagger}_{t_s}(n_s) |0 \rangle
\end{equation}
interpreted as corresponding to the proposition that one quantum
characterised by $n_i$ is present at time $t_i$ for $i = 1 \ldots s$. The
corresponding 
energy is given by the eigenvalues, e.g.
\begin{equation}
H_t |t_1, n_1 \rangle = \delta(t - t_1) \left(\frac{n_1 \pi}{L(t)} + E^0(t)
\right) |t_1, n_1 \rangle
\end{equation}
To be precise the Hamiltonian, the annihilation and creation operators as
well as the eigenvalues can be rigorously defined only with respect to
 smearing. The smearing functions for $a$ $a^{\dagger}$ and the eigenstates
 are elements of ${\cal E}$, or since $n$ is a discrete quantity with 
elements of $L^2(R)$, e.g.
can meaningfully define
\begin{eqnarray}
a_n(z) = \int dt z(t) a_t(n) \\
|n, \psi \rangle = \int \psi(t) |n,t \rangle
\end{eqnarray}
The smeared Hamiltonian $H_{\xi} = \int dt \xi(t) H_t$ is a well defined
operator on ${\cal F}$. This can be seen, in direct analogy with \cite{ILSS}
 by considering the automorphism
\begin{equation}
e^{-iH_{\xi}} a_t(n) e^{i H_{\xi}} = e^{-i \frac{n \pi}{L(t)} \xi(t)} a_t(n)
\end{equation}
which implies that $H_{\xi}$ can be rigorously defined on ${\cal F}$ by its
action on coherent states. If we denote by $\Gamma(A)$ the operator on 
${\cal F} = \exp {\cal E}$ defined as
\begin{equation}
\Gamma (A) | \exp z(.) \rangle = | \exp Az (.) \rangle
\end{equation}
where $A$ an operator on ${\cal E}$. Also defining  $d \Gamma (A)$ by 
\begin{eqnarray}
\Gamma (e^{A}) = e^{ d \Gamma(A)}
\end{eqnarray}
we can verify that 
\begin{equation}
H_{\xi} = i d \Gamma ( -i h_{\xi})
\end{equation}
with $h_{\xi}$ acting on ${\cal E}$ by
\begin{equation}
(h_{\xi} z)_n (t) = \frac{n \pi}{L(t)} \xi(t) z_n(t)
\end{equation}
Equation (3.30) implies that $H_{\xi}$ is actually definable for all
measurable $\xi(t)$ \cite{ILSS}.

\subsection{The decoherence functional}
We have now all the necessary information to compute the decoherence
functional:
\subsubsection{The Heisenberg operator} 
To construct the Heisenberg operator we start from equation (2.14) and have
to take into account the time-dependence of the Hamiltonian 
and the continuity of time. \\ \\
Let us first assume that time $t$ takes values in the interval $(t_0,t_f)$.
This can be taken as $(- \infty, \infty)$ if one wishes, but
 then square integrability forces that the boundary Hilbert spaces are just
 ${\bf C}$ and contain only the vacuum state. 
\par
Now, concerning the time - dependence  it  is straightforward to verify
that each operator $U_t$ at a single time Hilbert space ${\cal H}_{t}$ ought
be implemented by
\begin{equation}
U_t | \exp w \rangle_{{\cal H}_{t}} = |\exp  \hat{u}_t w \rangle_{{\cal H}_t}
\end{equation}
with $\hat{u}_t$ a unitary operator acting on ${\cal H}_t$ by
\begin{equation}
(\hat{u}_t w)_n = \exp \left( -i n\pi \int_{t_0}^t ds/L(s) \right) w_n
\end{equation}
Let us now pass to the continuous case. Using equation (3.15) we have
\begin{eqnarray}
{\cal U} |\exp w(.) \rangle_{{\cal F}} = \otimes_t U_t | \exp w_t
\rangle_{{\cal H}_t}  \nonumber \\
= \otimes_t |\exp \hat{u}_t w_t \rangle_{{\cal H}_t} = | \exp Uw
(.)\rangle_{{\cal F}}
\end{eqnarray}
where 
\begin{equation}
(U w)_n(t) = \exp \left( -i n \pi \int_{t_0}^t ds/L(s) \right) w_n (t)
\end{equation}
Hence \begin{equation}
{\cal U} = \Gamma(U)
\end{equation}
\subsubsection{The Schr\"odinger operator}
 In the construction of the decoherence functional the operator ${\cal S}$
 contains the information of the natural way the single-time
 Hilbert spaces are welded together. It involves a natural isomorphism
 between different - time Hilbert spaces. 
A natural expression for such an isomorphism would be to consider the map
\begin{equation}
I(t+a,t) | \exp w \rangle_{{\cal H}_{t}} = | \exp w \rangle_{{\cal H}_{t+a}}
\end{equation}
which is essentially the unitary operator connecting $b_t$ with  $b_{t+a}$
\footnote{Note that in this the index $t$ is simply  labeling the
 Hilbert space, $b_t$ meaning an annihilation operator in the single time
 Hilbert space ${\cal H}_t$. We are using the letter $b$ to distinguish from
 the operator valued distribution $a_t$ defined previously  as acting on
 ${\cal F}$.}.
But this would fail to perform the transformation $\phi_t \rightarrow
\phi_{t+a}$, which is deemed  essential, if this would be to connect the
 representations of the canonical groups at different single time Hilbert
 spaces. The identification of ${\cal H}_t$ at different $t$ ought to be
 given in terms of the physical variables of the theory, rather than the
 annihilation and creation operators, related to the non - invariant notion
 of particle.
\par
This being our concrete physical principle for the identification of time
translations, we can easily establish that these ought to be implemented
  on the $b$ and $b^{\dagger}$ as
\begin{eqnarray}
 b_t(n)   \rightarrow \frac{1}{2} \left( \sqrt{\frac{L(t+a)}{L(t)}} +
 \sqrt{\frac{L(t)}{L(t+a)}} \right) b_{t+a}(n) \nonumber \\
+ \frac{1}{2} \left( \sqrt{\frac{L(t+a)}{L(t)}} -\sqrt{\frac{L(t)}{L(t+a)}}
\right) b^{\dagger}_{t+a}(n)
\end{eqnarray} 
But can this transformation be unitarily implemented?
Before addressing this, let us recall a number of useful objects appearing in
the Fock space construction.  Let $V_t^*$ denote the complex conjugate of the
Hilbert space $V_t$ 
\footnote{Recall that $V_t$ by virtue of the representation of the history
group can be viewed as a real vector space with a specific complex
structure.}, with the act of complex conjugation defining an anti - linear
isomorphism  $C: 
V_t \rightarrow V_t^*$. Clearly $C^{-1} = - C^{\dagger}$.
\par
 An element $X(w)$ of the Lie algebra of the canonical group at time $t$ is
 parametrised by elements $w$ of $V_t$ as $X_t(w) = b(Cw) + b^{\dagger}(w)$.
 In terms of its matrix elements we have
\begin{equation}
\langle \exp z| X_t(w) | \exp \phi \rangle = \left[ (w,\phi)_{V_t} +
(z,w)_{V_t} \right] \langle \exp z | \exp \phi \rangle
\end{equation}
With this notation, the unitary operator $I(t+a,t)$ (which we shall denote
$I_a$ ) ought to act
\begin{equation}
I_a^{\dagger} \left( b_t(Cw) + b_t^{\dagger}(w) \right)I_a = b_{t+a}(CA_a w +
B_aw) + b_{t+a}^{\dagger}( A_aw + C^{-1}B_aw) 
\end{equation}
in terms of the  two operators
\begin{eqnarray}
A_a : V_t &\rightarrow& V_{t+a} \nonumber \\
(A_aw)_n &=& \frac{1}{2} \left( \sqrt{\frac{L(t+a)}{L(t)}} +
\sqrt{\frac{L(t)}{L(t+a)}} \right) w_n
\\
B_a : V_t & \rightarrow& V_{t+a}^* \nonumber \\
(B_aw)_n &=& \frac{1}{2} \left( \sqrt{\frac{L(t+a)}{L(t)}} -
\sqrt{\frac{L(t)}{L(t+a)}} \right) w_n
\end{eqnarray}
These operators can be easily checked to satisfy the Bogolubov identities
identities
\begin{eqnarray}
A_a^{\dagger}A_a - B_a^{\dagger}B_a = 1 \\
A_a^{\dagger} \bar{B_a} = B_a^{\dagger} \bar{A}_a
\end{eqnarray}
where we have denoted $\bar{B} = C^{-1}B$. Also we shall denote $\bar{A} = C
A C^{-1}$. 
\par
Unfortunately $I_a$ turns out not to be unitarily implementable. 
The necessary condition for this would
\begin{equation}
Tr (K_a^{\dagger} K_a) < \infty
\end{equation}
where  $ K_a : V_t \rightarrow V_{t+a}^* $ is defined by $K_a =
\bar{B_a}\bar{A_a}^{-1}$ 
In our case  $K^{\dagger}_a K_a$ is
\begin{equation}
(K^{\dagger}Kw)_n = \left(\frac{L(t+a)-L(t)}{L(t+a)+L(t)}\right)^2  w_n
\end{equation}
This is proportional to unity and hence its trace diverges.
\par
This is not surprising, it is the reason for the well-known inability to
define an instantaneous Hamiltonian in any field theory in non-static
 spacetimes\footnote{ We should  point out that time translation does not
 constitute in an sense a symmetry of the system 
( See Ref. \cite{Schr} for the implementation of symmetries in histories
theories.). This also implies that the action 
operator introduced in \cite{Sav} as the generator of physical time
transformations does not exist. This is 
 a consequence of the fact that such systems as quantum field theory in non -
 static backgrounds are effectively ``open''. The full theory ought to 
include the interaction with gravitational field, in which case (as it is
true  at the classical level) the Schr\"odinger time could be considered as 
homogeneous.  In this  case the  action operator might be expected to exist.
}.   
But this problem is not as serious   in a histories theory  as in a canonical
scheme. The intertwiners between the different representations of the
canonical group may not exist,
 but we should recall that the physical object in any history theory is the
 decoherence functional. The presence of continuous time (the important 
difference between this scheme and standard quantum mechanics) implies a
smearing in time of all physical quantities and will turn out to be crucial 
in our ability to provide a well defined decoherence functional.
\par
Let us start by regularising the operator $I$: we shall  consider an
ultraviolet integer cut-off $N$ in the field modes. The regularised operator
 $I^{(N)}$ is then well defined, and can be readily found to have the matrix
 elements 
\begin{eqnarray}
{}_{{\cal H}_{t+a}}\langle \exp w|I_a^{(N)}|\exp z \rangle_{{\cal H}_t} =
[\det( 1 -  \bar{K}_a K_a)]^{-1/2} \nonumber \\
 \exp \left( - \frac{1}{2}(w,K^{(N)}_aw) - \frac{1}{2} (\bar{K}^{(N)}_a z,z) +
 (A^{-1}w,z) \right)
\end{eqnarray}
At the limit $N \rightarrow \infty$ the Fredholm determinant diverges. 
The important point, for what  follows, is that when $a$ is taken to be
infinitesimal, that is, equal to $\delta t$ the $N$-dependence  appears
solely 
in the terms of order $(\delta t)^2$, and higher. It is easy to check that
for small small $\delta t$ , $K_a$ is proportional to $\delta t$ and since
 in the divergent determinant $K_a$ appears squared the lowest divergent
 contribution is of order $(\delta t)^2$. This point is of primary importance 
for the construction of a cut-off independent decoherence functional and as a
mathematical fact it is not restricted to our particular model.
\par
Let us now begin our construction of the Schr\"odinger operator. We shall
start by considering its discrete version using the regularised intertwiner.
 Let us assume then a discrete n-time history with propositions in the
 interval $[t_i,t_f]$ such that $t_i = t_0$ and $t_f = t_n$. Let the time
 interval between any two propositions be constant and equal to  $\delta t =
 (t_f - t_i)/n$.
 We have
\begin{eqnarray}
\langle \exp  w_{t_0}; w_{t_1}; \ldots ; w_{t_n} | {\cal S} | \exp z_{t_0};
z_{t_1};\ldots z_{t_n} \rangle \nonumber \\
= \langle \exp w_{t_0}|I^{(N)}_{\delta t}|\exp z_1 \rangle \langle \exp
w_{t_1}|I^{(N)}_{\delta t}|\exp z_{t_2} \rangle \ldots 
\langle \exp w_n | I^{(N)}_{t_n - t_0}|\exp
z_1 \rangle \nonumber \\ 
= \langle \exp w_{t_f}|I^{(N)}_{t_n - t_0}|\exp z(t_0) \rangle 
 \exp \left( \sum_{k=0}^{n-1} 
\left((w_k, z_{k+1}) \right. \right. \nonumber \\
\left. \left.
- \frac{\dot{L}(t_k)}{4 L(t_k)} ((w_{t_k},Cw_{t_k} ) + 
(C^{\dagger} z_{t_{k+1}},z_{t_{k+1}})) \delta t \right) 
+ O[(\delta t)^2]\right)
\end{eqnarray}
Now at the limit of $\delta t \rightarrow \infty$ the term in the exponential
in (3.48) becomes a Stieljes integral: hence at the continuous 
 limit one can meaningfully write
\begin{eqnarray}
\langle \exp w(.) | {\cal S} | \exp z(.) \rangle = 
\langle \exp w(t_f)|I^{ (N)}_T|\exp z(t_0) \rangle \hspace{3cm}\nonumber \\
\times \exp \left(  \sum_{n=1}^{\infty} 
\left( w^*_n(t_f) z_n(t_f) + \int_{t_0}^{t_f}ds  (w^*_n(s) z_n(s) + w^*_n(s)
\dot{z}_n(s) \right. \right. \nonumber \\
\left. \left. - \frac{\dot{L}}{4L}(s) [ w^*_n(s) w^*_n(s) + z_n(s) z_n(s)]
\right) \right) \nonumber \\ 
:= \langle \exp w(t_f)|I^{ (N)}_T|\exp z(t_0) \rangle e^{i A[w(.), z(.)] } 
 \end{eqnarray}
Hence the only divergent conribution to ${\cal S}$ comes from the operator
$I^{(N)\dagger} $ which appears in the boundary term.
 But it is the boundary term that is multiplied by the boundary operator
 ${\cal A}$ . To see how this multiplication is to be carried out let
 us return to equations (2.10) and (2.12). The indices $i$ and $j$ in
 equation (2.10) correspond respectively to the initial and final Hilbert
 space.
 Hence, when due to equation (2.12) the density matrix at the opposite
 boundary acts upon them one should introduce a factor of $I^{(N)}_T$.  
 Hence if one writes ${\cal A}$ in a coherent state basis on ${\cal H}_{t_0}
 \otimes \bar{{\cal H}}_{t_n} $ it should read
\begin{eqnarray}
 \langle \exp w_f|A^{u_fv_0}| \exp z_0 \rangle = \langle \exp w_f
 |(\rho_f)^{1/2} | \exp u_f \rangle \langle \exp v_0 | (\rho_0)^{1/2}
 I^{\dagger (N)}_T | \exp z_f \rangle \nonumber \\
\times [Tr(I^{(N)}_T \rho_f I^{\dagger(N)}_T \rho_0)]^{-1/2} \hspace{3cm}
\end{eqnarray}
where the presence of the $I^{(N)}$ in the denominator is to make the trace
well defined and $T = t_n - t_0$. 
Hence we can compute the map ${\cal AS}$ (properly speaking,the
multiplication is to be performed  in the discrete history version, but 
the $e^{iA}$ term is not affected anyway). 
Hence our result reads
\begin{eqnarray}
\langle \exp w(.) |{\cal S}{\cal A}^{u_fv_0} | \exp z(.) \rangle =
\langle \exp w(t_f) | (\rho_f)^{1/2} | \exp u_f \rangle  \langle \exp v_0
|(\rho_0)^{1/2}| \exp z(t_0) \rangle \nonumber \\
\times e^{iA[w(.),z(.)]}   [Tr(I^{(N)}_T \rho_f I^{\dagger(N)}_T
\rho_0)]^{-1/2}    \hspace{3cm}
\end{eqnarray}
There is still a $N$-dependence in the denominator, {\it but this cancels
out} when we restrict ourselves to the standard case of time-asymmetric
 histories with $\rho_f = 1$.
\par
Hence eventually, we have arrived at a well- defined, finite expression for
the decoherence functional for the time- asymmetric case. It is of
 the form (2.19) with the Heisenberg operator given by (3.36) and the map
 ${\cal SA}: {\cal F} \rightarrow {\cal H}_{t_i} \otimes \bar{{\cal
 H}}_{t_f}$
 given by the matrix elements
(3.51). 
\par
For completeness we give the following expression involving the trace over
${\cal H}_{t_0} \otimes \bar{{\cal H}}_{t_f}$
\begin{eqnarray}
  Tr_{{\cal H}_{t_0} \otimes \bar{{\cal H}}_{t_f}}({\cal SA})^{\dagger}
  ({\cal AS}) :{\cal F} \otimes \bar{{\cal F}} \otimes {\cal F} \otimes 
\bar{{\cal F}} & \rightarrow & {\bf C} \nonumber \\
|\exp w(.) \rangle \otimes \langle \exp z(.) | \otimes | \exp w'(.) \rangle
\otimes \langle \exp z'(.) | &\rightarrow& \nonumber \\
 \langle \exp w(t_f)| \exp z'(t_f) \rangle \langle \exp w'(t_0) |\rho_0| \exp
 z(t_0) \rangle    e^{ i A[w(.),z(.)] - i A^*[w'(.),z'(.)]}
\end{eqnarray}

Let us at this point address an important mathematical subtlety. The Gaussian
integral over the coherent states is essentially an integral of the Wiener
type 
defined primarily on skeleton paths (cylinder sets) and then by continuity
extending to the whole of the Hilbert space. As such one should be very
 careful when taking the continuum limit for the path. 
\par
This is in particular important for equation (3.49). There we have written a
term $\int dt  w^*(t)(1 + \frac{\partial}{\partial t})z(t)$
 as a limit of the discretised term $\sum_k w^*_k z_{k+1}$. It should be kept
 in mind that this is just a formal suggestive expression, so should 
not be taken literally. This should be more correctly  be written as 
\begin{equation}
\int dt w^*(t) (1 + \partial_-) w(t)
\end{equation}
Where $\partial_-$ is the {\it backwards} in  time Ito derivative defined by
the limit of the following matrix in skeleton paths
\begin{eqnarray}
\left( 
\begin{array}{clcl}
... & ...&...&...  \\ 
1 & -1  & 0 & ... \\
0 & 1& -1 &0    \\
0 & 0 &1 & -1  \\
\end{array}
\right) 
\end{eqnarray}
Also note that in ${\cal S}^{\dagger}$ the corresponding term is $ 1 -
\partial_+$ with $\partial_+ = - \partial_-^{\dagger}$ .
 For a large class of calculations this distinction might not be proved
 important, but whenever one encounters determinants and inverses,
 one should regularise and then take the continuum limit. This is the case
 for the calculations performed in the Appendix.
 For more details see for instance the reference \cite{Jan}.

\subsection{Some examples}
From our previous results it is easy to write down an explicit expression for
the decoherence functional evaluated for particular choices of
 history propositions. 
 \subsubsection{Coherent state histories and the path integral}
The projection operators 
\begin{eqnarray}
P_{w(.)} = | \exp w(.) \rangle \langle  \exp w(.) | /\langle \exp w| \exp w
\rangle
\end{eqnarray}
corresponds  to propositions about coherent state paths \cite{IL}. It is
straightforward to compute that
\begin{eqnarray}
d(w(.),z(.) ) =  \langle \exp w'(t_0) |\rho_0| \exp  z(t_0) \rangle
\langle \exp  w'(t_f)| \exp z'(t_f)
 \rangle \nonumber \\
\times
   e^{ i S[w(.)] - i S^*[z(.)]}
\end{eqnarray}
with $S$ the coherent state action
\begin{eqnarray}
iS[w(.)] = \sum_{n=1}^{\infty} \left( w^*_n w_n(t_f) - \int_{t_0}^{t_f} ds
\left( w^*_n(s) \dot{w}_n(s) - i \frac{n \pi}{L(s)} w^*_n(s) w_n(s) \right.
 \right. \\
\left. \left. + \frac{\dot {L}}{4L}(s)  [ w^*_n(s) w^*_n(s) + w_n(s) w_n(s)]
\right) \right.
\end{eqnarray}

The above formula has such a strong similarity to a coherent state path
integral expression for the decoherence functional, that we cannot help
wonder
 whether such an object is meaningful. 
\par
In some (not very precise) sense it is. If one wants to evaluate the time-
evolution kernel 
$ {}_{{\cal H}_{t_f}} \langle w_f; t_f|  w_i; t_i \rangle_{{\cal H}_{t_0}}$
one can proceed by the standard way by splitting the 
interval $[t_0,t_f]$ into intervals of width $\delta t$ and considering
evolution first by the intertwiner $I_{\delta t}$ and then by Hamiltonian
 evolution. It would be then easy to repeat the derivation that at the limit
 $\delta t \rightarrow 0$ the amplitude becomes $N$ - independent, reading 
 \begin{equation}
  {}_{{\cal H}_{t_f}} \langle w_f; t_f|  w_i; t_i \rangle_{{\cal H}_{t_0}} =
  \int Dw Dw^* e^{i S[w(.),w^*(.)]}
\end{equation}
with summation such that $w(t_0) = w_i$ and $w^*(t_f) = w^*_f$. But of course
this would be just a formal expression since ne cannot interchange
 the limit of $\delta t \rightarrow 0$ with the integrations. In addition, it
 would be rather awkward to have the standard quantum theory with 
changing Hilbert spaces. This construction is  natural only in a history
framework.
\par  
Still, it would be interesting to look for a rigorous definition of the path
integral (3.57). I am in particular referring to Klauder's algorithm 
\cite{Kla} of constructing the coherent state path integral through the use
of a metric on phase space, so that one can define a Wiener process upon it.
 It is a conjecture, worth investigating,
that such an object could be constructed from the introduction of a time -
dependent metric on phase space. The natural candidate would be the standard:
 the pullback of the projective Hilbert space metric to  phase space
\begin{eqnarray}
ds^2(t) = ||{\bf d} |qp;t \rangle||^2  - | \langle qp;t |{\bf d} |  qp; t
\rangle |^2 \nonumber \\
= \sum_{n=0}^{\infty} \left(\frac{n \pi}{2 L(t)} dq_n^2 + \frac{L(t)}{2 n
\pi} dp_n^2 \right) 
\end{eqnarray} 
where ${\bf d}$ is the exterior differentiation operator on phase space.

\subsubsection{Particle creation}
The best way to examine the effect of particle creation is through the
consideration of energy histories. That is we need to consider the value of
 the decoherence functional on coarse grained projectors in values of energy.
\par
The linearity of the field allows us to separately consider the effect on
each mode. Restricting ourselves to any mode labeled by $n$ we can easily
 verive that the most general projector onto energy eigenstattes is of the
 form
\begin{equation}
P = \sum_{r = 0}^{\infty} \int dt_1 \ldots dt_n \kappa_r(t_1,\ldots ,t_r)|
t_1 \ldots t_r \rangle \langle t_1 \ldots t_r |
\end{equation}  
where $\kappa_r$ are step functions elements of $L^2(R^r)$ and correspond to
smearing with respect to time of a proposition about $r$ quanta.
 Substituting these into equation (2.19) we could easily get an expression
 for the decoherence functional. Of particular interest is of  the case where
\begin{eqnarray}
\kappa_r(t_1, \ldots , t_r) = \chi_{\Delta}(t_1) \ldots \chi_{\Delta}(t_m)
\delta_{rm}
\end{eqnarray}
correponding to a proposition of a appearance of $m$ quanta (of the quantum
number n) within the time interval $\Delta$ ($\chi_{\Delta}$ stands
 for the characteristic function of this interval). 
 \par
It would be indeed cumbersome to conduct the full analysis of finding the
proper coarse graining of the energy histories that would allow us to
 identify consistent sets of energy histories and hence of the quasiclassical
 values of the total energy in the cavity.
\par
Still, we can make a number of qualitative statements solely through the
analysis of the logic of this construction. Essentially, the continuous
 history approach has enabled us to define a quantum theory by considering
 propositions of quantities smeared in time. Hence a proposition about energy
 is meaningful not when defined sharply at a moment, but rather as a
 proposition about the time - averaged number of excitations in a time
 interval of width $\Delta t$. Hence when evaluating the time average energy
 in an interval $\Delta t$ one has an effective high  energy cut -off
 at the mode number
\begin{equation}
N \simeq \frac{L}{\Delta t}
\end{equation}
From equation (3.42) one can estimate that (at the classical limit) the total
number of particles of mode $n$ created in an interval $\Delta t$
 is of the order of
 $(\dot{L} \Delta t/L)^2$, so that the average energy should peak around the
 value (ignore constants of order one)
\begin{equation}
\bar{E} \simeq \sum_{n = 1}^N \left(\frac{\dot{L} \Delta t}{L}\right)^2
\frac{n}{L} \simeq \frac{\dot{L}^2 (\Delta t)^2}{L^3}
 N^2 \simeq \frac{\dot{L}^2}{L}
\end{equation}
We can also estimate an optimal degree of coarse graining that minimises the
energy uncertainy  $\Delta E$. This ought to have two contributions ,
 the fully quantum ( proportional to $(\Delta t)^{-1})$ and a statistical one
 associated with time averaging. The latter will be essentially
\begin{equation} 
(\Delta E)_{av} \simeq \frac{\dot{L}^2 (\Delta t)^2}{L^3} N \simeq \frac{
\dot{L}^2 \Delta t }{L^2}
\end{equation}
hence the total energy fluctuation will behave as
\begin{equation}
\Delta E \simeq (\Delta t)^{-1} + \frac{ \dot{L}^2 \Delta t }{L^2}
\end{equation}
which is minimised at $\Delta t \simeq \frac{L}{\dot{L}}$ 
 Hence $\Delta E = \frac{\dot{L}}{L} $ is a minimum degree of energy coarse
 graining that will possibly lead to consistency of time - averaged
 energy histories . Since $\Delta E >> \bar{E}$ the classical picture we
 would get would be of large {\it classical } fluctuations around the minimum
 value $E^{0}(t)$. The fluctuations should be adequately described by a noise
 term of amplitude $\dot{L}/L$ when sampling in times larger than
 $L/\dot{L}$.

\subsubsection{ CTP generating functional and n-point functions}
The decoherence functional written in (2.19) is defined on pairs of
projectors of the history Hilbert space. As such it can be extended by
continuity
 to act on all bounded operators there. A particular instance is of course
 the smeared field operators. It can be readily checked that if
 for $P_{\alpha'} $ we substitute $\phi(X) \phi(Y)$ and for $P_{\alpha}$ the
 unity in (2.19), the value of the decoherence functional is nothing
 but the Feynman propagator for the field, i.e. the expectation value of the
 time - ordered product of two fields.
\par
This is an important point, because this means that from the decoherence
functional one can read objects appearing in the standard canonical
 quantum mechanical treatment. More generally if for $P_{\alpha'}$ we
 substitute the product $\phi(X_1) \ldots \phi(X_n)$ and for 
$P_{\alpha}$ $\phi(Y_1) \ldots \phi(Y_m)$ (smeared of course with suitable
test functions) we obtain in an obvious notation
\begin{equation}
d(\phi(Y_1) \ldots \phi(Y_m),\phi(X_1) \ldots \phi(X_n)) = G^{(n,m)}
(X_1,\ldots X_n,Y_1,\ldots Y_m)
\end{equation}
where $G^{(n,m)} $ are the Schwinger - Keldysh close- time - path (CTP)
correlation functions : for $m =0$ they are the time - ordered and for $n = 0
$ 
the anti -time ordered correlation functions. The corresponding generating
functional is then readily defined as
\begin{equation}
Z_{CTP}[J_+,J_-] = d(e^{-i \hat{\phi}(J_-)}, e^{i \hat{\phi}(J_+)})
\end{equation}
in terms of the smearing functions $J_+(t,x) $ and $J_-(t,x)$ interpreted as
external sources. Or still we can generalise to a phase space 
closed - time -path (PSCTP) generating functional
\begin{equation}
Z_{PSCTP}[(f,g)_+, (f,g)_-] = d(U^{\dagger}(f_-,g_-), U(f_+,g_+))
\end{equation}
in terms of the generators  $U(f,g)$ of the history group \footnote{The
relation of the generating functional to the decoherence functional is
 reminiscent of the definition of the generating functional of a stochastic
 process as a Fourier transform of the stochastic probability measure. 
In the quantum case we have a double Fourier transform which reflects the
fact that the decoherence functional represents a ``quantum measure'' 
on phase space \cite{Sor,An1}.}. 
\par
These equations can be easily verified to give the correct results in the
static case $\dot{L} = 0$ . They
  also provide well defined and finite objects in the general time -
  dependent case. This means in particular that the $n - point$ functions of
  this theory
 are meaningful distributions. Again we should stress the importance of
 smearing over time. In the canonical quantisation scheme, time is not
 treated
 in the same footing as the spatial variables, it is not a variable with
 respect to which one actually smears. The n-pt functions are then strictly
 speaking smooth functions of their $t$ - arguments. Here, smearing over $t$
 implies that the $n$-point functions are of non - trivial  distributional
 character with respect to all their variables.
\par
Let us consider the easiest case, where time runs in the full real axis and
the initial state is the vacuum. Then the time - ordered two - point
 function is
(recall $Tr {\cal S} = 1$).
\begin{equation}
G^{(2,0)} (t,x;t',x') = Tr \left({\cal U} \phi_t(x) \phi_{t'}(x') {\cal
U}^{\dagger}  {\cal S}^{\dagger} \right) 
\end{equation}
and can actually be computed (see the Appendix) as
\begin{eqnarray}
 G^{(2,0)} (t,x;t',x') = \sum_n \frac{ L(t) + L(t')}{[L(t) L(t')]^{1/2}}
 \frac{1}{4n \pi}
\sin \left( n \pi x \right) \sin (n \pi x') \left[ e^{-i n \pi |F(t) -
F(t')|} \right] \nonumber \\
 + \frac{1}{L(t)}\delta(t-t') \delta(x -x') \hspace{3cm}
 \end{eqnarray}
where $F(t) = \int_{- \infty}^{t} ds /L(s)$.
This is to be compared with the expression used by  Davies and Fulling
\cite{Ful}. 
\par
We should note here that in a histories framewok there is no direct physical interpretation for the $n$ -point functions. 
It is difficult to view them as expectation values of time - ordered products of the fields, simply because we need 
no assume an ensemble of systems - hence expectation value is a vague or even meaningless concept. This of course relates to the 
old problem of what meaning  quantum mechanical probability has when talking about a single system. In any case, the 
$n$ -point functions arise naturally as temporal correlation functions  from which the CTP generating  functional (and consequently the 
decoherence functional) is constructed. 
Indeed we would expect an analogue of Wightman 's reconstruction theorem \cite{SW} to hold in the histories 
version of quantum field theory. By that we mean  that the knowledge of the hierarchy of all CTP correlation functions 
(satisfying certain spacetime symmetry requirements ) should uniquely determine the histories Hilbert space ${\cal F}$,
 the decoherence functional and a representation of the group of spacetime symmetries on ${\cal F}$. 
\par
We should also remark that the study of the short distance  behaviour of the $n$ - point functions will
 enable us to determine whether they are of the Hadamard form \cite{Wald,FSW} (we do expect that for vacuum initial states 
taken as in the example we calculated above, but not for generic initial conditions). This will enable a direct 
comparison between the histories quantisation and the $C^*$ - algebraic framework \footnote{In particular the 
GNS construction of 
a Hilbert space where the $C^*$ - algebra is represented based on the choice of Hadamard vacua.} for the 
description of quantum fields in curved spacetime 
(see \cite{Wald} and references therein).

\section{Conclusions}
We have seen the construction  of a well defined, finite quantum theory
describing an 1+1 field in a time dependent cavity . This has been written
 in a continuous - time  histories form, using recently developed ideas and
 techniques and in this context being based on the used of smeared in time
 observables to ensure finiteness of our objects. It is in this sense
 important that we have been able to {\it a posteriori} justify our
 construction  by relating it to the CTP formalism . 
\par
We should remind again the reader the two important principles - one
mathematical one physical - entering our construction. The first is the use 
of the history group and the requirement of existence of an instantaneous
Hamiltonian as posited in \cite{IL,ILSS}. The second is the appearance 
 of two distinct notions of time - transformations as identified in
 \cite{Sav}. Their  separation  ought to be reflected 
in the probability assignment, hence in the decoherence functional. In a
concrete sense, our result points implies the construction of quantum
theories
in a histories scheme, that cannot be satisfactorily defined in a canonical
way \footnote{Another example, this time in classical systems, where constructions using histories
yield better results than the corresponding canonical ones is to be found in \cite{Koul}.}.
\par
From the perspective of the current paper the next step will be to apply
these principles  to provide a {\it unique} algorithm for constructing 
  a generic quantum field theory in curved spacetime. Our approach does
  straightforwardly apply in this case (  for spacetimes with compact Cauchy
 surfaces and in the absence of zero modes); the important unaddressed issue
 is whether the resulting history theory is dependent or not in the choice
 of the time variable: essentially whether the linear transformations
 associated with the change of foliation can be implemented  by {\it proper} Bogolubov
 transformations. This is a difficult problem both at the technical and conceptual level  
  and we hope to
address it in a future work.
\par
We should point out, that quantisation based on the history group has a
number of advantages over its canonical counterpart:
\\ \\
1. It has a much smaller degree of arbitrariness in the choice of the
representation. If, as we believe, changes of foliation turn out to be
unitarily implementable , then there will be no ambiguity at all in the quantisation
 algorithm.
\\
2. It allows a real - time description of field observables, rather than
focusing on evaluating the $S$ - matrix between in and out vacua. In
particular, global quantities (such as time - averaged total energy) 
can be unambiguously identified and we are allowed to make predictions or
assign probabilities 
about their values \footnote{ Even at the current state of the theory, where the 
general covariance of the scheme remains unproven, our results should be relevant to the study
of cosmological models, where a preferred foliation is always assumed. The
strength of this approach lies in the possibility of providing a unique
choice for the correlation functions of the field and as such should
facilitate treatment of issues like matter backreaction to the geometry,
especially in approaches that employ the CTP formalism (for a sampling of  recent work see \cite{Back}).}. 
\\ 
3. As a quantisation algorithm it is strictly local. That means, if we
restrict ourselves in a thin spacetime slice the theory is  constructed from 
the knowledge of the causal structure (${\cal S}$), the dynamics (${\cal U}$)
and the initial data (${\cal A}$). While the canonical approach necessitates 
a choice of positive frequency solutions and most physical criteria for such
a choice necessitate a knowledge of the behaviour of classical solutions 
at all
 times.
\\
\par
If our construction passes the test of unitarily incorporating changes of
foliation, this would be an impetus for further generalisation. Eventually, 
we would like to treat fields in spacetimes that are not globally hyperbolic:
ideally the ones appearing in the black hole evaporation process
 \cite{Har2}.
\section{Aknowledgements}
I am primarily indebted to K. Savvidou for insisting on the identification of
two laws of time evolution in history theories ; also for many
 discussions. I would also like  to thank C. Isham for comments on the first
 version of the paper and earlier discussions. Also I thank A. Roura for 
discerning an error in the first version of this paper.
\par
The explicit connection of the history formalism with the CTP was motivated
by discussions with B. L. Hu. Earlier discussions with E. Verdaguer 
and A. Roura have helped sharpen my understanding of this issue.
\par
The research was initiated at the University of Barcelona under the grant
96SGR-48 of the Generalitat de Catalunya and completed at the University of Maryland under
 NSF grant
PHY98-00967.

\begin{appendix}
\section{Computing the two - point function}
From equation (3.67) and the definitions (3.19) we see that the Fourier
transform of the   time - ordered two - point function  $G_{nm}^{(2)}(t,t')$
  is  by 
\begin{equation}
G_{nm}^{(2)} (t,t') = \delta_{nm} G_n (t,t')
\end{equation}
with
\begin{equation}
G_n(t,t') = \frac{[(L(t) L(t')]^{1/2}}{2 n \pi} Tr \left( {\cal U} (a_{tn}
a_{t'n} + a^{\dagger}_{tn} a^{\dagger}_{t'n} + a_{tn} a^{\dagger}_{t'n} 
+ a^{\dagger}_{tn} a_{t'n} ) {\cal U}^{\dagger} {\cal S}^{\dagger} \right)
\end{equation}
From (3.35) we can see that
\begin{equation}
{\cal U} b^{\dagger}_{nt} b_{t'n} {\cal U}^{\dagger} = e^{-in \pi [F(t) -
F(t')]}
\end{equation}
with $F(t) = \int_{-\infty}^t ds / L(s)$.
Similar expressions hold for the other terms in (A.2).
Hence we are left to the calculation of objects of the form ${\cal
S}^{\dagger} b_t b_{t'}$ etc. These are best computed by differentiation of
the generating functional
\begin{eqnarray}
A[F,F^*] = \int Dw Dw^* \exp \left( \int ds [- w^*(s) \partial_+ w(s) \right.
\nonumber \\
\left. + \frac{\dot{L}}{4L}(s) (w^*(s) w^*(s) + w(s) w(s)) + F^*(s)  w(s) +
w^*(s) F(s) ] \right)
\end{eqnarray}
Let us compute first the case $\dot{L} = 0$. In that case it is a standard
result that 
\begin{equation}
A[F,F^*] = \exp (\int ds ds'  F^*(s) \Theta_+(s,s')  F(s'))
\end{equation}
Here $\Theta_+$ is the inverse of $\partial_+$ and has matrix elements 
\begin{equation}
\Theta_+^{-1}(t,t') = \theta(t - t')
\end{equation}
Note the importance of having identified the operator as $\partial_+$. For
$\partial_-$ we would have
\begin{equation}
\Theta_-^{-1}(t,t') = - \theta(t' - t)
\end{equation}
In the general case where $\dot{L} \neq 0$ the integral is performed by
splitting $w$ into its real and imaginary parts. That is we define
\begin{eqnarray}
x(t) = (\frac{a}{2L(t)})^{1/2} (w + w^*) \\
y(t) =i ( \frac{L(t)}{a})^{1/2} (w^* - w)
\end{eqnarray}
in terms of some arbitrary positive real number $a$. Then the integral
becomes 
of the form
\begin{equation}
\int Dx Dy \exp \left(-  \frac{1}{2} X^TAY + K^T X \right) = \exp \left( K^T
A^{-1} K \right)
\end{equation}
with
\begin{eqnarray}
A = \left( 
\begin{array}{cl}
\frac{a}{L(t)}\partial_+ & -i \partial_+ \\
i \partial_+ & \frac{L(t)}{a} \partial_+
\end{array}
\right) 
\end{eqnarray}
\begin{eqnarray}
K = \left(
         \begin{array}{c}
         (\frac{a}{2L(t)})^{1/2} (F + F^*) \\
          i ( \frac{L(t)}{a})^{1/2} (F^* - F)
\end{array}
\right)
\end{eqnarray}
The inverse of $A$ is just 
\begin{eqnarray}
A^{-1}  =\frac{1}{2} \left( 
\begin{array}{cl}
\Theta_+ L/a &  i \Theta_+ \\
- i \Theta_+ & a \Theta_+ L^{-1}
\end{array}
\right) 
\end{eqnarray}
It is therefore easy to compute 
\begin{eqnarray}
A[F,F^*] = \exp \left( \frac{1}{2} \int ds ds' F^*(s) \left( \sqrt{\frac
{L(s)}{L(s'}} +    \sqrt{\frac{L(s')}{L(s)}}\right) \theta(s-s')  F(s')
\right)
\end{eqnarray}
leading to  
\begin{eqnarray}
G_n(t,t') = \frac{[(L(t) L(t')]^{1/2}}{4 n \pi} \left( \sqrt{\frac
{L(t)}{L(t'}} +    \sqrt{\frac{L(t')}{L(t)}}\right) e^{- i n \pi |F(t) -
F(t')|}
 + \delta(t - t')
\end{eqnarray}
from which (3.68) follows.
\end{appendix}

\end{document}